\documentclass[aps,pra,preprint,showpacs,tightenlines,nofootinbib,eqsecnum]{revtex4}
\usepackage{graphicx,bm,amsmath}

\begin{document}



\def\nid{\noindent}
\def\beq{
  \begin{equation}}
\def\eeq{\end{equation}}
\def\eeql#1{\label{#1}\end{equation}}
\def\bea{
  \begin{eqnarray}}
\def\eea{\end{eqnarray}}
\def\eeal#1{\label{#1} \end{eqnarray}}

\def\half{\mbox{\small $\frac{1}{2}$}}
\def\quarter{\mbox{\small $\frac{1}{4}$}}
\def\sfrac#1#2{\mbox{\small $\frac{#1}{#2}$}}
\def\bphi{\bm{\phi}}
\def\bpsi{\bm{\psi}}
\def\fu{\underline{f}}
\def\gu{\underline{g}}
\def\hu{\underline{h}}
\def\f{\bm{f}}
\def\g{\bm{g}}
\def\h{\bm{h}}
\def\p{\bm{p}}
\def\phiu{\underline{\phi}}
\def\phihu{\underline{ {\hat\phi} } }
\def\psiu{\underline{\psi}}
\def\psihu{\underline{ {\hat\psi} } }
\def\ot{\! \otimes \!}
\def\om{\omega}
\def\Om{\Omega}
\def\ep{\epsilon}
\def\H{{\cal H}}
\def\J{\chi}
\def\bS{\bm{S}}
\def\Nj{( \f_j , \f_j )}
\def\Nk{( \f_k , \f_k )}
\def\Re{\mathop\mathrm{Re}\nolimits}
\def\Im{\mathop\mathrm{Im}\nolimits}
\newcommand{\phd}{{\vphantom{\dagger}}}

\def\bdot{\mbox{\boldmath $\cdot$}}

\def\scalefig#1{\def\epsfsize##1##2{#1##1}}


\title{Eigenvector Expansion and Petermann Factor for Ohmically Damped Oscillators}

\author{Alec \surname{Maassen van den Brink}}\email{alec@dwavesys.com}
\author{K.~Young}
\author{M.H.~Yung}\thanks{Presently at Dept.\ of Physics, Univ.\ of Illinois at Urbana-Champaign, Urbana, IL 61801-3080, USA}
\affiliation{Physics Department, The Chinese University of Hong
Kong, Hong Kong, China}

\date{\today}
\pacs{42.55.Ah, 42.60.Da, 05.20.-y, 05.40.-a}

\begin{abstract}
Correlation functions $C(t) \sim \langle \phi(t) \phi(0) \rangle$ in ohmically damped systems such as coupled harmonic oscillators or optical resonators can be expressed as a single sum over modes~$j$ (which are not power-orthogonal), with each term multiplied by the Petermann factor (PF) $C_j$, leading to ``excess noise" when $|C_j| > 1$. It is shown that $|C_j| > 1$ is common rather than exceptional, that $|C_j|$ can be large even for weak damping, and that the PF appears in other processes as well: for example, a time-independent perturbation ${\sim} \ep$ leads to a frequency shift ${\sim} \ep C_j$. The coalescence of $J$ (${>}1$) eigenvectors gives rise to a critical point, which exhibits ``giant excess noise" ($C_j \rightarrow \infty$). At critical points, the divergent parts of $J$ contributions to $C(t)$ cancel, while time-independent perturbations lead to non-analytic shifts ${\sim} \ep^{1/J}$.
\end{abstract}

\maketitle

\section{Introduction}
\label{sect:intro}

In open optical cavities, spontaneous emission rates and thermal noise in a mode $j$ are modified by a factor $C_j$ compared to the conservative case, which can paradoxically exceed unity, giving rise to ``excess noise". This possibility was first pointed out by Petermann~\cite{petermann79} for gain-guided semiconductor lasers, and $C_j$ has come to be known as the Petermann factor (PF). This prediction was initially controversial~\cite{controv}, until Haus and Kawakami~\cite{haus85} pointed out the importance of correlation with other modes. Siegman~\cite{siegman89} showed that excess noise is present in all open-sided laser resonators or optical lens guides. Cheng and Siegman~\cite{cheng} suggested a quantization scheme for non-orthogonal modes and derived the excess-noise factor quantum-mechanically. New~\cite{new} provided examples where the PF can be large. Berry~\cite{berry} has given a more general account, stressing the importance of degeneracies. The subject continues to be debated in the optics community~\cite{aiello02}.

The recent literature is largely confined to optical systems, with focus on specific examples and modes; this paper will however emphasize the generality of the PF in three ways. (a)~The PF applies to many systems, not just optical resonators and not only in the quantum domain. (b)~For each system the PF is relevant to a range of processes. (c)~It is the rule rather than the exception for the PF to exceed unity.

First, the issues are relevant for any ohmically damped system (see Section~\ref{sect:concl} for further generalizations). In linear conservative systems, thermal correlation functions $C(t)\sim\langle\phi(t)\phi(0)\rangle$ can be written as a single sum over eigenstates or normal modes $j$. The same holds in ohmically damped systems, but their eigenstates, called \emph{quasinormal modes}, have complex frequencies $\om_j$ and are not power-orthogonal. We show, within this general context, that each term in this expansion is multiplied by $C_j$, and that this factor can exceed unity. The broader context exhibits the concepts clearly, bypasses the complexities of particular modes in optical resonators, allows simple finite-dimensional examples to be studied, and shows that the PF is essentially classical. Incidentally, there is much recent interest in gravitational-wave detectors---a damped classical system, for which naive use of power-orthogonal modes fails, on account of inhomogeneous losses~\cite{grav}. The method developed here is a powerful and convenient tool for the modal analysis of the noise in such systems.

Second, for any given system, the PF occurs in many processes and quantities, e.g., in the response to time-independent perturbations. The ubiquity of this factor stems from the mathematical structure of eigenvector expansions for damped systems~\cite{dissa,dissb}, where orthogonality is defined not with respect to the familiar inner product $\langle \bpsi | \bphi \rangle$, but with respect to a symmetric bilinear map $(\bpsi, \bphi)$. The PF is essentially the ratio between these two mathematical constructs---the inner product which defines length, and the bilinear map which defines orthogonality.

Section~\ref{sect:lin} briefly recapitulates the linear-space structure and expansion in eigenvectors~$\f_j$, introducing the bilinear map. The PF
$C_j \propto 1/ (\f_j , \f_j)$ is then defined for each mode $j$. It is shown that $|C_j|\ge1$ (a)~for each underdamped mode ($\Re\om_j\neq0$) and (b)~when averaged over all modes (the latter statement being relevant when many modes are overdamped). This serves to correct the impression that ``excess noise" may be a rare phenomenon.

Section~\ref{sect:cor} derives the expansion for correlation functions~$C(t)$. Instead of the \emph{double} sums over modes often seen in the literature (e.g., Ref.~\cite{seig}), with off-diagonal terms sometimes referred to as mode--mode coupling, exact sum rules (\ref{eq:sumrule}) here simplify $C(t)$ to a \emph{single} sum~(\ref{eq:cor06}), involving the PF~$C_j$ but otherwise similar to the conservative limit.

Then, Section~\ref{sect:pert} shows that the familiar Rayleigh--Schr\"odinger perturbation theory applies to the \emph{complex} eigenvalues, provided eigenvectors are normalized by $(\f_j,\f_j)$. Thus a time-independent perturbation ${\sim}\ep$ leads to shifts ${\sim}\ep C_j$, i.e., possibly ``excess" response. Examples of ``giant excess noise" ($C_j\rightarrow\infty$) are easily constructed in simple systems with few degrees of freedom.

This invites a number of questions: for example, what happens if $C_j \rightarrow \infty$, or equivalently $(\f_j , \f_j)\rightarrow 0$?  Section~\ref{sect:corcrit} shows that this happens if and only if $J>1$ eigenvectors merge at a \emph{critical point}, where the divergent parts of the $J$ terms cancel, leaving a finite result; the divergent PF is then manifested as perturbations ${\sim}\ep$ causing large and non-analytic frequency shifts ${\sim}\ep^{1/J}$. A model is studied in which a critical point off the imaginary frequency axis is split. This model is rich yet solvable, and illustrates many interesting features, especially in the weak-damping limit where there is an interplay between two small parameters: $\delta = \left|\Im\om_j\right|$ and the splitting $\lambda$ between two nearly-critical modes. It is shown that the PF goes as $C_j \sim \delta/\lambda$, and that there can be prominent time-domain effects. Concluding remarks and a sketch of further generalizations are given in Section~\ref{sect:concl}.

The general formalism allows contact with two well-known results. First, critical damping (in the elementary sense of that word) of one single oscillator is precisely a case of ``giant excess noise"---thus helping to demystify the PF. Second, $(\f_j,\f_j)$, and thus also the PF, is related to the Zeldovich normalizing factor for quasinormal modes~\cite{zel}. The present formalism also reveals the classical, rather than quantum, origin of the PF.

This paper emphasizes the role played by the PF, especially a number of issues not addressed previously: the relevance of the PF to many different phenomena, the connection among these, the fact that the PF typically (rather than exceptionally) exceeds unity, the divergence at critical points, its cancellation in physical quantities, its manifestation as non-analytic shifts in time-independent perturbation theory around a critical point, and the interplay between weak damping and near-degeneracy. As a result of the general formalism, correlation functions (and related quantities) are expressed as \emph{single} sums over modes, a significant simplification.

\section{Linear-space structure}
\label{sect:lin}

\subsection{Formalism}
\label{formalism}

The PF and related issues mentioned above can be transparently studied in damped harmonic systems. Consider $N$ coupled classical linear oscillators labelled as $\alpha = 1, \ldots, N$, with coordinates $\phi= (\phi(1),\ldots,\phi(N))^{\rm T}$, described by\footnote{The equation of motion can be given a quantum pedigree if one starts with a bath which is then eliminated from the equations of motion or path integral.}
\beq
  \left( M d_t^2 + \Gamma d_t + K \right)  \phi = 0 \quad ,
\eeql{eq:eqmot1}
where $M(\alpha,\beta) = m_{\alpha} \delta( \alpha,\beta)$ is a diagonal mass matrix (but allowing for coordinate transformations, any symmetric positive-definite matrix), $\Gamma\ge0$ is a symmetric damping matrix and $K>0$ is a symmetric matrix of force constants. Classical fields are included if $\alpha$ is turned into a continuous position variable $x$ (Appendix~\ref{app:a}). Gain media can be described by relaxing $\Gamma \ge 0$ while some dispersive media  can be described in a restricted frequency range by relaxing $M>0$. Incidentally, $\phi_M \equiv M^{1/2} \phi$ satisfies (\ref{eq:eqmot1}) with $M \mapsto I$, $\Gamma \mapsto \Gamma_M
\equiv M^{-1/2}\Gamma M^{-1/2}$ and $K \mapsto K_M \equiv M^{-1/2}
K M^{-1/2}$, a unit-mass representation that is sometimes useful.

In the presence of dissipation it is convenient to write the dynamics in first-order form, by introducing the momentum $\hat{\phi} = ( \hat{\phi} (1) , \ldots , \hat{\phi}(N) )^{\rm T}$ and the phase-space state $\bphi = (\phi,\hat{\phi} )^{\rm T}$. Then the evolution takes the Schr\"odinger-like form
\beq
  d_t \bphi = -i \H \bphi \quad ,\qquad
  \H =i\begin{pmatrix}0  & M^{-1} \\ -K & -\Gamma M^{-1}\end{pmatrix} \quad .
\eeql{eq:defh}
With the compensating factors of $-i$ and $i$ in (\ref{eq:defh}) there is an analogy with quantum mechanics, and more importantly the eigenvalues of $\H$ are the frequencies.

In phase space, the inner product $ \langle \bpsi  | \bphi \rangle
\equiv  \psi^{\dagger} \phi + \hat{\psi}^{\dagger} \hat{\phi} $
satisfies $\langle \bphi  | \bphi  \rangle \ge 0$. Unfortunately,
$ \langle \bpsi  | \H \bphi  \rangle \neq \langle \bphi  | \H
\bpsi  \rangle^*$, which renders it useless for projections or
orthogonality.  Instead, we use a bilinear map~\cite{dissa}, which will be key to the PF:
\beq
  ( \bpsi , \bphi ) = ( \bphi , \bpsi ) \equiv i \bigl(
  \psi^{\rm T} \hat{\phi} + \hat{\psi}^{\rm T} \phi + \psi^{\rm T}
  \Gamma \phi \bigr)\quad .
\eeql{eq:blmap}
Note that $M$ does not appear in the bilinear map, as is readily seen by transforming from the unit-mass representation. This bilinear map is constructed to ensure the key property
\beq
  (\bpsi,\H\bphi) = (\bphi,\H\bpsi) \quad ,
\eeql{eq:symh}
analogous to self-adjointness. In proving (\ref{eq:symh}), the dissipative term in $\H$ cancels the last term in (\ref{eq:blmap}). The diagonal entries $(\bphi,\bphi)$ are not positive definite, not even necessarily real; they may also be small (or even zero)---a central concern of this paper. Equivalently to (\ref{eq:blmap}), one could have introduced a duality map $D$, so that $(\bpsi,\bphi)=\langle D\bpsi|\bphi\rangle$. Since $D$ maps right eigenvectors to left ones, the eigenexpansion (\ref{eq:dyn1}) is then seen to be a biorthogonal expansion. This is the language used in, e.g., Refs.~\cite{siegman89,new,berry}; see Ref.~\cite{dissa} for details.

Eigenvalues are the roots of  the characteristic function $\J(\om) \equiv \det (\H - \om)$, and eigenvectors are defined by $\H \f_j = \om_j\f_j$. For an eigenvector~$\f_j$, (\ref{eq:defh}) implies the time dependence $e^{-i\om_j t}$, so the coordinates and momenta are related: $\hat{f}_j= -i\om_j M f_j $. With dissipation, the eigenvalues are complex. For $\Gamma \ge 0$ and $K > 0$, $\Im \om_j \le 0$.\footnote{\label{damp}This is readily demonstrated as follows. Multiply the defining equation by $f_j(\alpha)^*$ and sum over~$\alpha$. Define $m = f_j^{\dagger} M f_j^\phd > 0$, $2 m \gamma = f_j^{\dagger} \Gamma f_j^\phd \ge 0$ and $k = f_j^{\dagger} K f_j^\phd  > 0$, giving $ -\om_j^2 m -2i\om_j m \gamma + k = 0$ familiar from the case of a single oscillator.} Upon conjugation, it follows that $-\om_j^{*}$ is also an eigenvalue, with eigenvector $ \propto \f_j^* $. Thus, except for $\Re  \om_j = 0$, eigenvalues are paired, ensuring that $\phi$ is real even though each eigenvector is complex. The property (\ref{eq:symh}) is the analog of self-adjointness and, as usual, leads to $(\f_j,\f_k)=0$ if $\om_j \neq \om_k$, henceforth called orthogonality. This condition is extended to level crossings [see below~(\ref{eq:ineq02})] by taking suitable linear combinations.

\newpage
Since $\H$ is not self-adjoint, the eigenvectors may be incomplete, but only on a set of measure zero in parameter space, to be called critical points for reasons that will become apparent. Until Section~\ref{sect:corcrit}, we assume that there are $2N$ eigenvectors forming a complete basis, so the dynamics given the initial condition $\bphi(t{=}0) = \bphi$ is readily solved:\footnote{One curious consequence of the expansion (\ref{eq:dyn1}) is the following. Suppose we want to excite the system with a given energy, say an initial condition $\bphi$ with $\langle\bphi|\bphi\rangle=1$, to maximize the amplitude of mode~$j$. It follows that the optimal choice is not $\bphi=\f_j$ but the left eigenvector $\bphi=\f^j\equiv D\f_j$, even though the latter will excite other modes besides $j$ as well. This is known as adjoint coupling; see, e.g., Ref.~\cite{adj-coupl} and references therein.}
\beq
  \bphi(t) = \sum_j \frac{ (\f_j , \bphi) } { (\f_j , \f_j ) }
  \, \f_j \, e^{-i\om_j t} \quad .
\eeql{eq:dyn1}
For $t=0$, this becomes a resolution of the identity. In terms of
$N \times N$ blocks,
\begin{align}
  0 &= \sum_j f_j {\otimes} f_j / \Nj \quad, \notag\\
  \displaybreak[0]
  I &= \sum_j \om_j (Mf_j) {\otimes} f_j / \Nj \quad, \notag\\
  \displaybreak[0]
  0 &= \sum_j f_j {\otimes} (\Gamma f_j) / \Nj \quad, \notag\\
  \displaybreak[0]
  0 &= \sum_j [ \om_j^2 (Mf_j) {\otimes} (Mf_j) + i\om_j (Mf_j)
  {\otimes} (\Gamma f_j) ] / \Nj \quad,\label{eq:sumrule}
\end{align}
where $a {\otimes} b$ stands for the matrix with elements
$a(\alpha) b(\beta)$. The denominator $(\f_j , \f_j)$ may be
small, which is at the heart of this paper.

The norm in coordinate space $N(\phi) = \phi^{\dagger} M \phi$ is a good measure of length, but the diagonal bilinear map $(\bphi , \bphi )$ is not (in part because it can vanish). The relationship between the two is of interest especially for an eigenvector $\f_j$, for which the momentum is not independent. We are therefore led to compare
\beq
  (\f_j , \f_j ) = f_j^{\rm T} \left( 2\om_j M + i \Gamma\right) f_j
\eeql{eq:norm1}
with
\beq
  N_j \equiv N(f_j) =  f_j^{\dagger} M f_j^\phd
\eeql{eq:norm2}
(which is different from $\langle \f_j | \f_j \rangle$ even for $M=I$). In the limit of zero dissipation, $(\f_j,\f_j)=2\om_j N_j$, which motivates the definition of the PF\footnote{The definition is arbitrary up to any reasonable factor that reduces to unity in the conservative limit. For example, $2\om_j$
can be replaced by its absolute value.}
\beq
  C_j \equiv  \frac{2 \om_j N_j}{ ( \f_j , \f_j) } \quad .
\eeql{eq:pet1}
Interestingly, the length [cf.~the norm (\ref{eq:norm2})] and the projection [cf.~the bilinear map (\ref{eq:blmap})] do not relate to the same inner product; in a sense $C_j$ measures the difference between the two. Several physical quantities depend on~$|C_j|$. However, since $N_j \propto f_j^* f_j^{\phantom{*}}$ whereas $(\f_j, \f_j)\propto f_j^2$, the phase of $C_j$ is a matter of convention.

To show that $|C_j|>1$ is common, take the trace of the second sum rule in (\ref{eq:sumrule}); this gives $2N =\sum_j \sigma_j C_j$, where
\beq
  \sigma_j = \frac{f_j^{\rm T} M f_j^\phd }{ f_j^{\dagger} M f_j^\phd }
\eeql{eq:sigma1}
is bounded by unity (for the usual case $M>0$); thus the average value of $|C_j|$ is  $\ge 1$.  For underdamped modes ($\Re\om_j \neq 0$), one can prove a stronger statement: from (\ref{eq:eqmot1}) in the frequency domain one has $-\om_j^2 m-2i\om_j m\gamma + k = 0$, where $m$, $\gamma$ and $k$ are defined in footnote~\ref{damp}. Then from (\ref{eq:norm1})  we find $(\f_j , \f_j )  = \om_j f_j^{\rm T} M f_j^\phd + \om_j^{-1} f_j^{\rm T} K f_j^\phd$. The magnitude of each term is increased if $f_j^{\rm T}$ is replaced by $f_j^{\dagger}$, and using the definitions of $m,\gamma, k$,
\beq
  | (\f_j , \f_j ) | \le | \om_j |^{-1} \left( |\om_j|^2 m + k
  \right) f_j^{\dagger} f_j^\phd \quad .
\eeql{eq:ineq02}
But for an underdamped mode, $| \om_j |^2 = k / m$, hence the RHS becomes $2 |\om_j| f_j^{\dagger} M f_j^\phd$. It then follows that $|C_j| \ge 1$ for every underdamped mode.  (Thus the statement about the average value is relevant only when many modes are overdamped.) These results imply that ``excess noise" [cf.\ below (\ref{eq:cor07})] is common rather than rare.

The formalism so far relies on two assumptions: (a) the eigenvectors are complete, and (b)~$(\f_j , \f_j) \neq 0$ for all $j$.  These two conditions are related. To  see that (a) implies (b), suppose that at a critical point an eigenvector is lost because of merging, say $\f_k \rightarrow \f_j$.  Then, $(\f_j ,\f_j) = \lim \, (\f_j , \f_k) = 0$. Conversely, suppose $(\f_j ,\f_j) = 0$.  Then if the eigenvectors are complete, $\f_j$ ($\neq0$) would be orthogonal to every vector $\bpsi$, a contradiction. If all eigenvalues are distinct, the $2N$ eigenvectors must be linearly independent. Thus, the eigenvectors can only be incomplete if $\J(\om)$ has a root of order $J>1$. When $J$ roots merge (\emph{degeneracy}) as a parameter is tuned, there are two possibilities: either there are still $J$ linearly independent eigenvectors (\emph{level crossing}), or the eigenvectors merge as well (\emph{criticality}). (The non-generic case where some but not all of the $J$ eigenvectors merge will be ignored except for Example 4 below.)  Incompleteness occurs only at criticality. We shall see that criticality is \emph{more} generic than level crossing (footnote~12 in Ref.~\cite{dissa})---somewhat surprising since for conservative systems, level crossing is allowed whereas criticality is not.

\subsection{Examples}

Because of its perhaps unusual properties, let us give several examples of the dissipative eigenexpansion and the classical PF, especially the possibility of it being large.

\emph{Example 1}. Consider just $N=1$ oscillator, with $M = 1$, $K = k$ and $\Gamma = 2\gamma$. The eigenvalue equation $\J(\om)=0$ leads to $ \om = \om_{\pm}=\pm\Omega-i\gamma$, where $\Omega=\sqrt{k-\gamma^2}$. The eigenvectors are $\f_{\pm} = A_{\pm} ( 1 ,  -i\om_{\pm} )^{\rm T}$, with the bilinear maps $( \f_{+} , \f_{-} ) = 0$, $(\f_{\pm},\f_{\pm}) = \pm 2 A_{\pm}^2 \Omega$, and the norm $N_{\pm} = |A_{\pm}|^2$. Choosing a convenient phase, we find
\beq
  C_{\pm} = \frac{ \Om \mp i\gamma }{\Om} \quad .
\eeql{eq:ex1-06}
We note that $C_{\pm} = 1$ if $\gamma=0$; in the underdamped regime $|C_{\pm}| \ge 1$, while in the overdamped regime, one of $|C_{\pm}|$ exceeds unity. At the critical point $k = k_* = \gamma^2$, the two eigenvalues merge: $\om_{+} = \om_{-} = -i \gamma$; the two eigenvectors also merge, leaving only one eigenvector in the 2-dimensional space; and the diagonal bilinear maps vanish while $C_{\pm}\sim\Omega^{-1}$ diverge.

\emph{Example 2}. Let $N=2$, $M=I$ and
\beq
  K = \begin{pmatrix} k_{11} & k_{12} \\ k_{12} & k_{22} \end{pmatrix} \quad ,
  \qquad \Gamma = 2 \begin{pmatrix} \gamma_{11} & \gamma_{12} \\
  \gamma_{12} & \gamma_{22} \end{pmatrix} \quad .
\eeql{eq:exb01}
With the freedom to rotate coordinates, without loss of generality assume $\gamma_{12}=0$ and consider the one-parameter family $k_{11} = k_{22} = 4$, $k_{12} = -2$, $\gamma_{11} =2\gamma$, $\gamma_{22} = \gamma$. For small $\gamma$, there are two pairs of underdamped modes. One pair goes critical at $\gamma_{*1} = 0.8599$, and the other pair at $\gamma_{*2} =
2.1031$, beyond which all modes are overdamped and eigenvalues purely imaginary.  Except at the two critical points, the eigenvectors are complete, and all bilinear maps etc.\ can be evaluated explicitly.

In these examples of critical damping (in the elementary sense), a pair of conjugate eigenvalues $\om_j$ and $\om_{-j}= -\om_j^*$ merge on the imaginary axis.  This occurs with codimension $= 1$ in parameter space, and is the most generic class of the merging of eigenvectors---which we therefore refer to as criticality in general; the relevant subspace is called a Jordan block.

\emph{Example 3}.  Critical points off the imaginary axis occur with codimension $> 1$ in parameter space. Consider two second-order blocks at $\om = \pm1 -i \delta$: $\det (\H-\om) =(\om{-}1{+}i\delta)^2(\om{+}1{+}i\delta)^2$.  The choice $\gamma_{12}=\gamma_{22}=0$ results in the one-parameter family
\beq
  K = \begin{pmatrix} 1 + 5\delta^2 & 2\delta \sqrt{1+\delta^2} \\
  2\delta \sqrt{1+\delta^2} & 1+\delta^2 \end{pmatrix} \quad ,
  \qquad \Gamma =
  \begin{pmatrix} 4 \delta & 0 \\ 0 & 0 \end{pmatrix} \quad .
\eeql{eq:exc01}
A small perturbation will result in nearly degenerate modes. This is dealt with in Section~\ref{sect:corcrit}; the case where both $\delta$ and the mode splitting are small has a number of interesting properties.

\emph{Example 4}. To obtain a block where $J=4$ eigenvalues merge, we require $\J(\om)$ to have a 4th-order zero. Take $N=2$, $M=I$ and (\ref{eq:exb01}) and set $ \mbox{det}(\H{-}\om)=(\om{+}i)^4$ by a choice of scale. This gives four equations, leading to a 2-parameter family of solutions, of which a simple 1-parameter subset is: $k_{11} = k_{22} = \cosh(x)$, $k_{12} = \sinh(x)$, $\gamma_{11} = 1 + s_3$, $\gamma_{22} = 1-s_3$, $\gamma_{12} = s_2$, where $s_n =  \sinh^n(x/2)/\sinh(x)$. Non-negativity of $\Gamma$ requires $\cosh(x)\le3$. The eigenvector(s) are found from $ (K - \Gamma + I ) f = 0$, and generically there is only one solution, hence a $J=4$ block (cf.\ the end of Section~\ref{formalism}). However, exceptionally there can be more than one eigenvector, leading to the crossing of different blocks.  For a $2 \times 2$ system, this requires $K-\Gamma+I=0$, and happens only for $x=0$---the trivial case of two independent but identical oscillators, each generating a $J=2$ block at the critical point. Interestingly, one cannot produce crossing between $J=3$ and $J=1$ blocks with only two oscillators.

These non-generic examples are much easier to construct and analyze than for continuum models (such as optical cavities). In all cases, one verifies the divergence of the PF at a critical point, in line with Berry's observations~\cite{berry}. However, in the first three examples, with $J=2$, the PF goes as the inverse of the mode splitting, $C_j\sim(\Delta\om_j)^{-1}$; in Example~4, with $J=4$, one sees from (3.5), (3.6), and (3.9), all in Ref.~\cite{dissb}, that $C_j\sim(\Delta\om_j)^{1-J}=(\Delta\om_j)^{-3}$. Perturbations of these examples will be examined below.

\section{Correlation functions}
\label{sect:cor}

\subsection{Formalism}
\label{subsect:tn}

Many physical processes are related to thermal correlation functions. When the oscillators are placed in a bath at temperature $T$, (\ref{eq:eqmot1}) acquires on the RHS a noise term $\eta(\alpha,t)$, which satisfies the fluctuation--dissipation theorem
\beq
  \langle \tilde{\eta} (\alpha,\om) \tilde{\eta}(\beta, \om')
  \rangle =  4 \pi T  \delta(\om{+}\om')  \Gamma(\alpha, \beta)
  \quad ,
\eeql{eq:fd}
in units with $k_{\rm B} =1$, where $\langle \cdots \rangle$ denotes thermal average and $\tilde{\phantom{\om}}$ denotes Fourier transform.

Write the equation of motion in two-component form:
\beq
  (d_t + i\H) \, \bphi (t)  = \bS (t) = ( 0 , \eta(t))^{\rm T}\quad .
\eeql{eq:eqmotf2}
Using $ \bphi(t) = \sum_j a^j(t) \f_j $, we find $(d_t + i\om_j) \, a^j(t) =(\bS(t) , \f_j)/\Nj$, where from (\ref{eq:eqmotf2}) and (\ref{eq:blmap}), and henceforth adopting the summation convention for Greek indices, $(\bS (t) , \f_j) = i \eta(\alpha', t) f_j(\alpha') $. Upon Fourier transform,
\beq
  \tilde{a}^j(\om) = - \frac{1}{\om-\om_j} \, \frac{
  \tilde{\eta}(\alpha',\om) f_j(\alpha') }{\Nj} \quad ,
\eeql{eq:dynm4}
from which $\bphi(t)$ is obtained. Again $\Nj$ could lead to large response to noise.

Consider the correlation function $C(\alpha, \beta ; t)=\langle\phi(\alpha,t)\phi(\beta,0)\rangle$. Using (\ref{eq:dynm4}) gives
\beq
  \tilde{C}(\alpha,\beta;\om) = - 2T \sum_{jk}
  \frac{ f_j(\alpha) \, [ f_j(\alpha')
  \Gamma(\alpha',\beta')  f_k(\beta') ] \, f_k(\beta) }
  {(\om-\om_j) (\om + \om_k)\Nj \, \Nk}\quad .
\eeql{eq:cor01}
From the definition of the bilinear map,
\beq
  f_j(\alpha') \Gamma(\alpha',\beta') f_k(\beta')
  = -i \Nk \delta_{j,k} + i (\om_j {+} \om_k) f_j(\alpha')
  M(\alpha', \beta')f_k(\beta')
  \quad .
\eeql{eq:cor02}
When the second term of (\ref{eq:cor02}) is put into (\ref{eq:cor01}), there is a factor $(\om_j{+}\om_k)/(\om{-}\om_j)(\om{+}\om_k)=(\om{-}\om_j)^{-1} - (\om{+}\om_k)^{-1}$, resulting in one term without $\om_k$ and another without~$\om_j$. The former leads to a sum $ \sum_k f_k(\beta') f_k(\beta)/ \Nk = 0 $ by (\ref{eq:sumrule}); likewise the latter vanishes.

The remaining first term in (\ref{eq:cor02}) then leads to the central result, which, in contrast to analogous formulas in the literature (e.g., Ref.~\cite{seig}), involves a \emph{single} sum over modes:
\bea
  \tilde{C}(\alpha,\beta;\om)&=& 2iT \sum_j
  \frac{f_j(\alpha) f_j(\beta)} {(\om^2-\om_j^2)\Nj}\nonumber\\
  &=& 2iT \sum_j \frac{1}{2\om_j(\om^2-\om_j^2)} \left[
  \frac{f_j(\alpha) f_j(\beta)}{N_j} \right] \, C_j \quad .
\eeal{eq:cor06}
Then, Fourier transforming (\ref{eq:cor06}) and evaluating the residues gives
\beq
  C(\alpha, \beta; t) = \frac{T}{2} \sum_j \frac{1}{\om_j^2}
  \left[ \frac{f_j(\alpha) f_j(\beta) }{N_j} \right] C_j \,
  e^{-i\om_j t} \quad .
\eeql{eq:cor07}

The square bracket in (\ref{eq:cor06}) and (\ref{eq:cor07}) is normalized in that its trace with $M(\alpha, \beta)$ is bounded by unity; if a mode has negligible dissipation ($\om_j$ real) or is overdamped ($\om_j$~imaginary), then $f_j$ has a constant phase, and the trace has unit modulus. Thus $C_j$ appropriately expresses the relative contribution of each mode. The familiar conservative case is recovered by setting all $C_j=1$. Herein lies the paradox: the response per mode to thermal noise can be \emph{increased} by dissipation (``excess noise") and the PF exceeding unity is the rule rather than the exception; near criticality, some $C_j$'s even diverge (``giant excess noise").

\subsection{Example}
\label{subsect:corex}

Example 1 with $N=1$ already serves to demystify the PF and its
possible divergence.  In this case, $f_{\pm}^2 / N_{\pm} =1$ and
$C_{\pm} = \pm \om_{\pm}/\Om$.  Some arithmetic leads to
\beq
  C(t) = \frac{T}{k}
  \frac{\om_- e^{-i\om_+ t} - \om_+ e^{-i \om_- t} }
  { \om_- - \om_+ } \quad .
\eeql{eq:cor08}
Although each term has a large coefficient (``excess noise")
near criticality, the sum is not large. In particular, $C(t)$ is
manifestly finite at the critical point where $\om_+ - \om_-
\rightarrow 0$.

A much simpler derivation can be given for this trivial case. In general, $\phi(t{>}0) = \phi(0) G_1(t) + \dot{\phi}(0) G_2(t)+\phi_{\eta}(t)$, where $G_i$ are homogeneous solutions satisfying the initial conditions $G_1(0) = 1$, $\dot{G}_1(0) = 0$, $G_2(0) = 0$, $\dot{G}_2(0) = 1$, and $\phi_{\eta}(t)$ is an inhomogeneous solution caused by and therefore proportional to $\eta(t{>}0)$. The last term has zero correlator with $\phi(0)$ and $\dot{\phi}(0)$, while $\langle \phi(0)^2 \rangle = T/k$, $\langle\phi(0) \dot{\phi}(0) \rangle =0$, which then leads to $C(t) = (T/k) G_1(t)$, in agreement with (\ref{eq:cor08}).


\section{Perturbation Theory}
\label{sect:pert}

The PF also occurs in perturbation theory, which takes the familiar Rayleigh--Schr\"odinger form, everywhere replacing the usual inner product with the bilinear map (\ref{eq:blmap})~\cite{dissa}, provided that map itself is unperturbed. Thus, we only consider changes in $K$,\footnote{More generally, changes in $M$ are allowed as well.}
namely $\H=\H_0+\ep\Delta\H$, with
\beq
  \Delta \H = i \begin{pmatrix} 0 & 0 \\ -\Delta K & 0\end{pmatrix}
  \quad .
\eeql{eq:ndp02}
For example, the first-order frequency shift is
\beq
  \Delta\omega_j=\ep\,\frac{(\f_j,\Delta\H\f_j)}{\Nj} \quad .
\eeql{eq:ndp04}
Higher-order terms in analogy to the conservative case will not be displayed. We have thoroughly verified (\ref{eq:ndp04}) in examples; interestingly, it correctly gives $\Im\Delta\omega_j$ as well.

The connection with the PF can be made more explicit:
\beq
  ( \f_j , \Delta \H \f_j ) = f_j^{\rm T} \, \Delta K \,
  f_j ^{\phantom{*}}\equiv(\Delta K)_{jj}N_j\quad,
\eeql{eq:ndp05}
which is the obvious way to define the normalization-independent matrix element $(\Delta K)_{jj}$ (and analogously for off-diagonal elements).  Then
\beq
  \Delta (\om_j^2) = \ep (\Delta K)_{jj} C_j \quad ,
\eeql{eq:ndp05a}
noting that the ``natural" eigenvalue for second-order dynamics is $\om^2$ rather than $\om$.  Thus the shift due to a perturbation ${\sim}\ep$ is ${\sim}\ep C_j$---with the possibility of large shifts if $|C_j| \gg 1$, much in parallel with ``excess noise" as large response to thermal fluctuations. Example~1 shows this property explicitly in an elementary setting: $\om = -i\gamma \pm \sqrt{k-\gamma^2}$, so upon $k \mapsto k + \ep\Delta k$, we have $\Delta \om = (\ep/ 2) \Delta k/\sqrt{k-\gamma^2}$, with divergent shifts near criticality.

Finally we note that writing perturbation theory in terms of $\Nj$ [cf.~(\ref{eq:ndp04})] exhibits the formal analogy with conservative systems, whereas writing it in terms of $N_j$ and $C_j$ [cf.~(\ref{eq:ndp05a})] emphasizes the possibility of anomalously large shifts.

A result equivalent to (\ref{eq:ndp04}) was given long ago by Zeldovich~\cite{zel}, who used an integral expression [the analog of $\Nj$] to normalize the perturbation matrix element for outgoing waves. The normalizing factor, originally involving a regulator, was later given in a more convenient form and applied widely~\cite{pert}. The generalization to off-diagonal bilinear maps through a first-order formalism~\cite{twocomp,openwavermp} in fact motivates our (\ref{eq:blmap}).

\section{Criticality and Near-criticality}
\label{sect:corcrit}

The \emph{finite} correlation function $C(t)$ is a sum over mode contributions $\propto C_j$, and each $C_j$ can be large. This apparent paradox takes an extreme form \emph{at} criticality, where $C_j$ diverges. Section~\ref{subsect:corex} already gave an example where the divergent parts cancel; this section shows this in general. Indeed, a similar cancellation occurs in all physical quantities involving~$C_j$.

\subsection{Interpolating function}
\label{subsect:intf}

Consider only eigenvectors $\{ \f_j \}_{j=1,\ldots, J}$ which merge as a parameter $\lambda$ approaches a critical point $\lambda_*$ ($=0$ without loss of generality). Let the eigenvalues be $ \om_j = \om_* + \lambda \zeta_j \rightarrow \om_* $. (Note that $\lambda$ measures the changes in eigenvalues, and is not proportional to the size $\epsilon$ of any perturbation in ${\cal H}$.) The coefficients $\zeta_j = e^{2\pi i j/J}$ from (\ref{eq:zetadef}) below label the directions and relative rates at which the frequencies approach the limit. We will need that for $| m | \le J{-}1$,
\beq
  S_m \equiv \frac{1}{J} \sum_j \zeta_j^m = \delta_{m,0} \quad.
\eeql{eq:sumzeta}

Now introduce a minimal polynomial $\f(\om) = \sum_{n=0}^{J-1} \g_n \, (\om-\om_*)^n$ to interpolate the $J$ eigenvectors: $ \f_j = \f(\om_j )$. The expression $(i d_t - \H) [f(\om) e^{-i \om t}]$ (say evaluated at $t=0$) vanishes at $\om = \om_j$, so in the critical limit has a zero of order $J$ at $\om=\om_*$; but as a polynomial of order $J{-}1$ in $\om$, it must then vanish identically. This leads to $(\H - \om_*) \g_n = \g_{n-1}$, for $n=0, ..., J{-}1$, with the convention $\g_{-1}=0$.

With a suitable normalization (cf.~\cite{dissb}), we can further achieve
\beq
  (\g_n , \g_{n'} ) = \delta_{n+n' , J{-}1 } \quad ,
\eeql{eq:jborth}
so that the diagonal bilinear map becomes
\beq
  \Nj = \sum_{n,n'} (\g_n , \g_{n'} ) \, (\lambda
  \zeta_j)^{n+n'} = J (\lambda\zeta_j)^{J-1} \quad .
\eeql{eq:normcrit}
The $\g_n$ are defined at each small $\lambda$, and the implicit assumption is that there is a finite limit as $\lambda \rightarrow 0$. Thus $\{ \g_n \}$ is more convenient than $\{ \f_j \}$ near criticality; the limiting $\{\g_n \}$ is the Jordan normal basis and the subspace spanned is called a Jordan block~\cite{dissb}.

We again illustrate with Example 1. Take $k = \gamma^2-\lambda^2$, so that the eigenvalues are $\om_{\pm} = -i(\gamma \mp\lambda)$ with the eigenvectors given above (\ref{eq:ex1-06}). The critical point occurs at $\om_* =-i\gamma$, and $\zeta_{\pm}= \pm i$. The minimal polynomial
interpolation is
\bea
  \f(\om) &=& \frac{ \f_+ + \f_- }{2} + \frac{ \om -
  \om_*}{\om_+ - \om_*} \frac{\f_+ - \f_-}{2}\nonumber \\
  &=& \begin{pmatrix} 1 \\ -\gamma \end{pmatrix}
    + (\om - \om_*) \begin{pmatrix} 0 \\ -i\end{pmatrix}\quad.
\eeal{eq:inter5}
The coefficients $\g_n$ are independent of $\lambda$ as
expected and also satisfy (\ref{eq:jborth}). Another example of
near-criticality is given in the next section.

\subsection{Small denominators}
\label{subsect:small}

The problem of divergent PFs, or $( \f_j , \f_j ) \rightarrow 0$, can be studied more generally. Consider (\ref{eq:cor06}), use (\ref{eq:normcrit}) for $\Nj$, put $\om_j = \om_* + \lambda\zeta_j$, and expand in powers of $\lambda$, giving terms $\propto\sum_j (\lambda \zeta_j)^{-J+1+\ell} $, with $\ell \ge 0$.  But from (\ref{eq:sumzeta}), all the negative powers sum to zero.  This resolves the small-denominator problem, namely the paradox that as $\lambda \rightarrow 0$, the contribution of each mode diverges (``giant excess noise"). We do not spell out the remaining finite terms, which are in any event better expressed in terms of the Jordan normal basis~$\{\g_n\}$.

Although the divergent parts cancel in the sum for $C(\alpha,\beta;t)$, one could choose linear combinations that single out only one mode.  What happens to the cancellation in that case?  For simplicity suppose two modes $j=1, 2$ are close to criticality, and consider
\beq
  \langle(d_t{+}i\om_1)\phi(\alpha,t)\,\phi(\beta,0)\rangle\quad.
\eeql{eq:elim1}
The operator $d_t {+} i\om_1$ eliminates the $j{=}1$ contribution, through an extra factor $-i(\om_j{-}\om_1)$ in (\ref{eq:cor06}) and (\ref{eq:cor07}). However, the remaining $j{=}2$ term will then carry a factor $-i(\om_2{-}\om_1)$, which also vanishes at criticality. Thus projecting out one near-critical mode suppresses other such modes, so that again the physical result is not large.  Contrast the conservative case,
where projecting out one mode does not at the same time suppress the others.


\subsection{Perturbation around critical points}
\label{subsect:pcrit1}

Section~\ref{sect:pert} showed that large PFs are also manifested
in time-independent perturbation theory: a perturbation
${\sim}\ep$ produces a frequency shift ${\sim}\ep C_j$, with $C_j
\rightarrow \infty$ at a critical point. So what happens if a
system originally \emph{at} a critical point is perturbed?  In
this section, we show that the shifts become non-analytic, and
generically ${\sim} \ep^{1/J} \gg \ep$.

Consider a perturbation $\H = \H_0 + \ep \Delta \H$, in which $\H_0$ describes a system at a critical point $\om_*$, where a block of $J$ eigenvectors have merged. In powers of $\ep$, the characteristic polynomial $\J(\om)$ is (considering this block only)
\bea
  (-1)^J\J(\om)&=&\J_0(\om)+\ep\J_1(\om)+\ep^2\J_2(\om)+\cdots\nonumber\\
  &=&(\Delta\om)^J+\ep[\J_1(\om_*)+\Delta\om\J'_1(\om_*)+\cdots\,]
  + \ep^2 [ \J_2(\om_*) + \cdots \, ] + \cdots \quad ,
\eeal{eq:pertchar1}
where $\J_0(\om)$ has a $J$th-order root $\om_*$ and $\Delta\om = \om - \om_*$. Setting $\J(\om)=0$ gives, to leading order,
\bea
  \Delta \om &=& [-\J_1(\om_*)]^{1/J} \, \ep^{1/J} \, \zeta_j +\cdots
  \quad,\label{eq:pertchar3}\\
  \zeta_j &=& e^{2\pi i j/J}
\eeal{eq:zetadef}
for $j = 1, \ldots, J$. Thus (a)~the shifts go as $\ep^{1/J}$ (non-analytic in $\ep$ and $\gg |\ep|$); (b)~the eigenvalues split into $J$ different ones, shifting in equiangular directions, all at the same rate; (c)~the directions of splitting for $\ep <0$ bisect those for $\ep > 0$. These features are already contained in Example~1: for $k = k_* + \epsilon$, the eigenvalues are $-i\gamma \pm \sqrt{\ep}$, approaching the critical point along the real (imaginary) direction for $\ep > 0$ ($\ep < 0$).

The myriad non-generic possibilities will not be exhausted. For example, if $\J_1(\om_*) = 0$ but $\J'_1(\om_*) \neq 0$, then to leading order,
\beq
  0 = \Delta \om \, [ (\Delta \om)^{J-1} + \ep \J'_1(\om_*) +\cdots ] \quad .
\eeql{eq:ngpert03}
Thus one state is unshifted to lowest order, while the other $J{-}1$ states split like a generic block of order $J{-}1$. [For $J=2$, the $\ep^2\J_2(\om_*)$ term is of the same order and must be retained as well.] A more systematic analysis involving the basis vectors as well is given elsewhere~\cite{dissb}.

Perturbations of Example~3 exhibit interesting features. Let $k_{12} \mapsto k_{12} + \epsilon$; the eigenvalues solved from $\chi(\om)=0$  bisect each other for opposite signs of $\ep$, resembling critical damping in the elementary sense---even though $\om_*$ is not purely imaginary. Their exact form reads
\bea
  \om_j &=& -i\delta \pm \sqrt{ 1 \pm \sqrt{ \ep ( 4\delta
  \sqrt{1+\delta^2} + \ep ) } }\label{eq:exc-p1} \\
  &\approx& -i\delta \pm \bigl( 1 \pm  \sqrt{\delta \ep + \ep^2/4 }
  \bigr) \quad ,
\eeal{eq:exc-p2}
where the two signs can be independently chosen, and where the last expression is valid for $|\ep|\ll1$, $\delta\ll1$ (but without assumption on their relative magnitudes). The separation between the two near-degenerate eigenvalues will be denoted as $2\lambda$ (the factor of 2 for consistency with Section~\ref{sect:corcrit}), and
\beq
  \lambda \approx \sqrt{\delta \ep + \ep^2 / 4} \quad ,
\eeql{eq:exc-p3}
which goes as $\ep^{1/2}$ for $|\ep| \ll \delta$ (the region for which perturbative results are valid), but as $\ep$ for $|\ep|\gg \delta$.  It is also straightforward to show that $|C_j|\propto 1/\lambda$ for all modes, provided $|\ep| \ll 1$, $\delta\ll 1$. Incidentally, if the system is perturbed by the opposite sign of $\ep$, then $\lambda \rightarrow i\lambda$ and the two modes split in the imaginary direction instead.


\subsection{Weak damping versus near-degeneracy}
\label{subsect:pcrit2}

PFs are nontrivial only in the presence of damping; one therefore expects effects proportional to $\left|\Im\om_j\right|\sim\delta$. PFs become large when modes are nearly degenerate; typically they go as $1/\lambda$ where $\lambda$ characterizes the separation between eigenvalues.  An interesting question is the interplay between $\delta$ and $\lambda$ for weakly damped and nearly degenerate modes. The above example serves to illustrate this regime, of interest for near-degenerate optical modes near gain threshold.

\begin{figure}[p]
\centering
\includegraphics{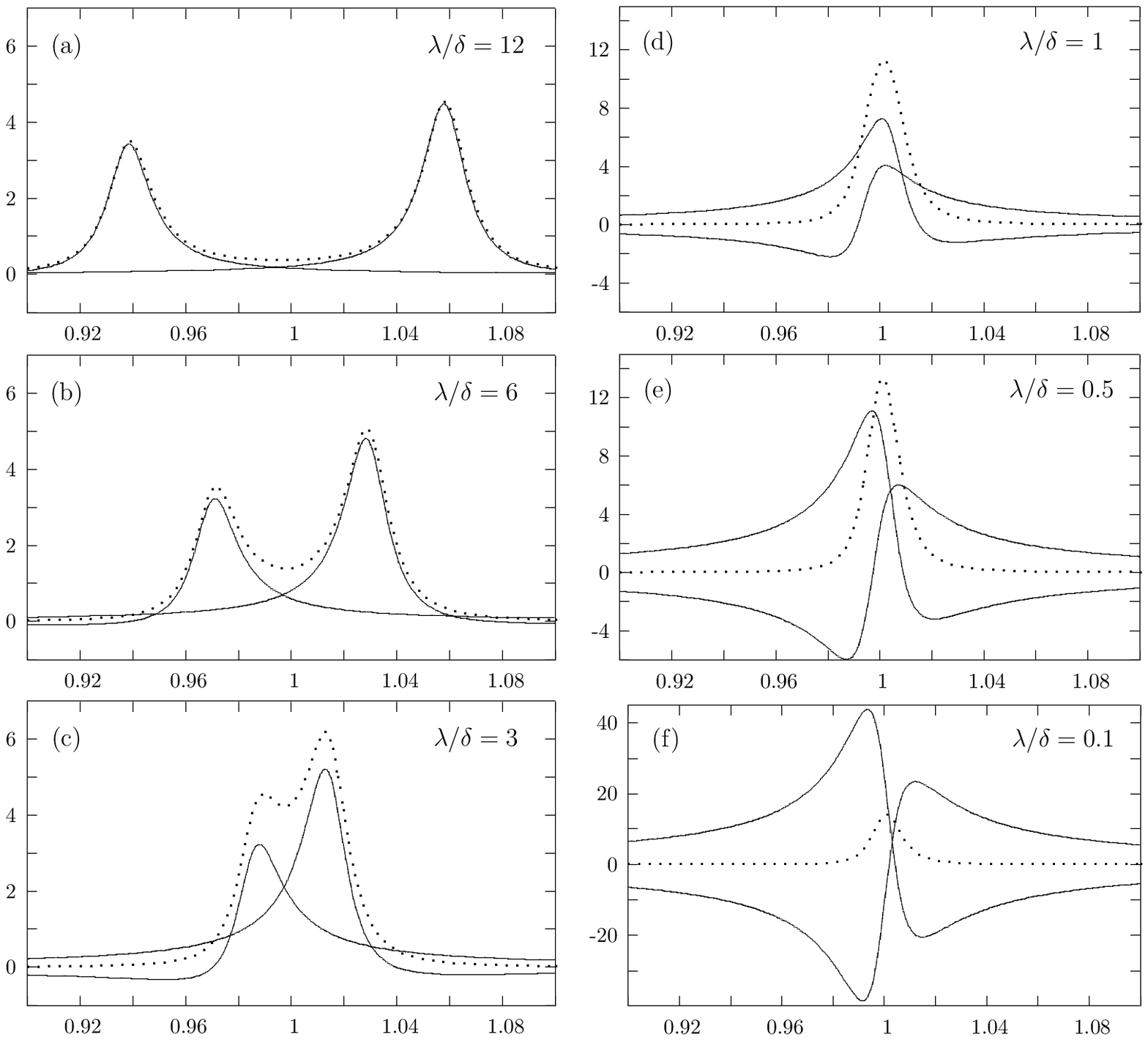}
\caption{Sequence of the correlator $\tilde{C}_X/100$ versus
$\om$ for two weakly damped and near-critical modes, for different
values of $\lambda / \delta$; see Section~\ref{subsect:pcrit2}.
The solid lines represent individual
mode contributions. The dotted line represents the sum.}
\label{fig3}
\end{figure}

For the model defined in Section~\ref{subsect:pcrit1} with eigenvalues given by the exact formula (\ref{eq:exc-p1}), consider the correlation function $\tilde{C}_X(\om) \equiv X(\alpha,\beta)\tilde{C}(\alpha, \beta; \om)$, where $X$ is any symmetric matrix in coordinate space. To be specific in the following we take $X=\bigl(\begin{smallmatrix} 1 & 1 \\ 1 & 7 \end{smallmatrix}\bigr)$. Figure~\ref{fig3} shows $\tilde{C}_X(\om)$ versus $\om$ for a fixed $\delta = 0.01$ and various values of $\lambda$. The contributions of each pair of modes $j=\pm1$ or $j=\pm 2$ (solid lines) are separate lorentzians for $\lambda/\delta  \gg 1$, merge when $\lambda / \delta \sim 1$, and diverge as $\lambda / \delta\rightarrow 0$.  (Actually, the contributions of $j=-1$ and $j=-2$ are negligible in the frequency range shown.)  However, the sum (broken line) remains finite even when $\lambda / \delta\rightarrow 0$. To understand the interplay between the two small parameters $\delta$ and~$\lambda$, it is convenient to consider the amplitudes $B_j$ of the respective lorentzians, defined by
\beq
  \tilde{C}_X(\om) = iT \sum_j \frac{B_j}{\om_j(\om^2-\om_j^2) } \quad ,
\eeql{eq:defbj}
where $B_j = \sigma_j^{\prime} C_j$, with
\beq
  \sigma_j^{\prime} = \frac{ f_j^{\rm T} X f_j } {
  f_j^{\dagger} M f_j } \quad ,
\eeql{eq:sigma2}
in analogy to (\ref{eq:sigma1}). We shall show in Appendix~\ref{app:b} that to leading order as $\lambda\rightarrow0$:
\beq
  C_{1,2} \approx \pm c \frac{\delta}{\lambda} \quad ,
\eeql{eq:clim}
for some $c=O(1)$, while for $\lambda \gg \delta$, $C_{1,2}\rightarrow 1$. Moreover, $\sigma_j^{\prime} = O(1)$. [However, $\sigma_j^{\prime} = O(\delta)$ for the special case $X=M$.] Thus we expect the following behaviour for $B_{1,2}$. (a)~They depend on $\lambda/\delta$ rather than on each separately. (b)~They go as $\pm \delta / \lambda$ (with the same coefficient) as $\lambda /\delta \rightarrow 0$. (c)~They approach (in general different) finite values as $\lambda /\delta \gg1$. These properties are verified by the numerical results in Figure~\ref{fig4} for $\Re B_j$ versus $\lambda / \delta$; the imaginary part is similar and not shown. The factor $\delta$ in (\ref{eq:clim}) can be explained heuristically: criticality has to disappear and become level crossing when $\delta \rightarrow 0$.

The form (\ref{eq:clim}) implies that there will be a significantly enhanced PF only when $\lambda \ll \delta$, but that is precisely the regime where the two lorentzians merge.  Thus, the effect of a very large PF will not be easily observable in equilibrium correlation functions.


\subsection{Late-time behaviour}
\label{subsect:noneq}

Infinite PFs are never physically observable. However, large PFs are observable in the time domain, even for weakly damped modes ($\left|\Re\om_j\right|\gg\left|\Im\om_j\right|$, as in the example just described). Let $j=\pm 1$ be the most weakly damped mode, with $\gamma = -\Im  \om_{\pm 1}$. Let $\gamma' > 0$ be a lower bound on $\Im  \om_{\pm 1} - \Im  \om_k$ for $k \neq \pm 1$. Then for times $t$ such that $\gamma' t \gg 1$, only the $j = \pm 1$ modes are relevant in the evolution (\ref{eq:dyn1}) and correlator~(\ref{eq:cor07}). The latter shows that $C_{\pm 1}$ is now directly measurable. This scenario can be achieved in the above example by taking $\ep<0$ (so that the modes split in the imaginary direction), and choosing $|\ep| \ll \delta \ll 1$ gives $\gamma'= 2 |\lambda| = 2\sqrt{\delta |\ep|}$, while $|C_{\pm 1}| \sim\delta /\lambda \sim  \sqrt{\delta / |\ep|} \gg 1$. [The slightly overdamped case of (\ref{eq:cor08}) is even simpler.] Thus, the large PF is observable, albeit as an algebraic enhancement of an exponentially small tail so that the total effect is still very small; cf.\ the discussion of (\ref{eq:elim1}) for fixed~$t$. In an optical cavity with one mode near gain threshold, $\gamma \approx 0$ and the surviving term is in fact constant in time.

\section{Conclusion}
\label{sect:concl}

In ohmically damped linear systems, including a broad class of optical resonators, the correlation function can be expressed as a sum over eigenvectors $f_j$, differing from the conservative case only through a Petermann factor (PF) $C_j$ in each term. In this paper we have demonstrated this in a broad context,  establishing many properties systematically, without reference to the details of cavity modes. For example, (complex) frequency shifts due to time-independent perturbations are $\propto C_j$. PFs are particularly interesting near critical points, where $C_j\rightarrow\infty$ when $J > 1$ eigenvectors merge; however, in the correlation function the divergent parts cancel, while in time-independent perturbation theory the shifts ${\sim}\ep C_j$ go over to ${\sim}\ep^{1/J}$.

In conservative systems one is used to Hilbert spaces, in which vectors have magnitudes (associated with diagonal inner products $\langle \bphi | \bphi \rangle$) and directions (directional cosines associated with off-diagonal inner products $\langle \bpsi| \bphi \rangle$); both refer to the \emph{same} inner product. In dissipative systems, lengths of eigenvectors are given by $N_j$ whereas projections are given by $( \bpsi , \bphi)$; the PF arises because these are \emph{different}. That is, the linear-space structure for dissipative systems~\cite{dissa} is significantly different from that for conservative systems. These require a first-order formalism, involving both coordinates and momenta (in the optics case, both the magnetic and electric fields), an ingredient previously missing in the literature. Interestingly, this mathematical structure leads to the PF, which is observable.

The present discussion for a finite number of linear classical oscillators is readily generalized. (a)~Many models of dispersion can be accommodated by enlarging the linear space,~or equivalently postulating hidden ohmic oscillators~\cite{cheung}. (b)~Turning $\alpha$ into a continuous variable $x$ gives continuum models, in which nearest-neighbor couplings (i.e., only between $\alpha$ and $\alpha{\pm}1$) turn into a second-order spatial derivative $\partial_x^2$; see Appendix~\ref{app:a}. Electromagnetic waves in optical resonators are then included. (c)~The variables $\bigl(\phi(\alpha),\hat{\phi}(\alpha)\bigr)$ can be promoted to operators satisfying $\bigl[\phi(\alpha,t)\,,\,\hat{\phi}(\beta,t)\bigr] = i \hbar\delta(\alpha,\beta)$. Eigenvector expansions remain formally unchanged, while equations of motion are modified only by the presence of quantum noise (which however needs to be handled with care). The coefficients $a^j$ and $(a^j)^*$ then become annihilation and creation operators. Certain two-point correlations are just the Feynman propagators, which can be used in a perturbative expansion for interacting (i.e., nonlinear) fields, much in the usual way. Interestingly, $\Delta^{jk} (t)\sim\langle a^j(t)^{\dagger}a^k(0)\rangle$ is not diagonal. In fact, the usual logical chain from plane electromagnetic waves to electromagnetic propagators to free photons to interacting photons can be simply repeated---with the difference that each mode in the expansion is now a quasinormal mode, with PF~$C_j$. This links our work to the extensive literature discussing the PF in quantum optics~\cite{quantum-PF}. Since the PF is purely a property of the modes, which follow from a classical wave equation in either case, it will play the same role, and have the same properties, in either the quantum or our classical~treatment.


\begin{acknowledgments}
This work builds upon a long collaboration with many colleagues: E.S.C. Ching, H.M. Lai, P.T. Leung, S.Y. Liu, W.M. Suen, C.P. Sun, S.S. Tong, and many others. KY thanks Richard Chang for discussions on optics in microdroplets as resonators, which initiated our interest in waves in open systems. Some earlier work on the linear-space structure for ohmically damped oscillators was carried out with S.C. Chee.
\end{acknowledgments}

\begin{appendix}
\section{Continuum model}
\label{app:a}

To establish the link between the oscillator models of the main text and cavity optics, consider a scalar model of electromagnetism:
\beq
  \mu \varepsilon( {\bf r} ) \frac{ \partial^2 \phi ( {\bf r} ,
  t) } {\partial t^2} + \mu \sigma ( {\bf r} ) \frac{ \partial  \phi
  ( {\bf r} , t) } {\partial t} - \nabla^2 \phi ( {\bf r} , t) = 0\quad ,
\eeql{eq:cont01}
where $\mu$ is the permeability of free space, $\varepsilon$ is the dielectric constant and $\sigma$ is the conductivity. Let the model be defined on $|x|\le a$, $0 \le y \le b$, $0 \le z \le c$, with $\phi = 0$ on the boundary.  Further assume that the system is uniform in $y$ and $z$: $\varepsilon = \varepsilon(x)$, $\sigma = \sigma(x)$; then the $y$ and $z$ dependence can be expressed as $ \phi \propto \sin (n_1\pi y/b) \sin (n_2 \pi z/c)$, and the wave equation reduces to the 1-d model
\beq
  \rho(x) \frac{ \partial^2 \phi(x,t) }{\partial t^2} +
  \Gamma(x) \frac{ \partial \phi(x,t) }{\partial t} + (K \phi) (x,t)
  = 0 \quad ,
\eeql{eq:cont03}
where $\rho = \mu \varepsilon$, $\Gamma = \mu \sigma$ and $K= q^2 - \partial_x^2$, with $q^2 = (n_1 \pi / b)^2 + (n_2 \pi/c)^2$; the boundary condition is $\phi(|x|{=}a,t)=0$. Our formulation studies the $t$-dependence, with eigenvectors evolving as $e^{-i\omega t}$; many works in the literature study mode propagation along the optic axis~$x$, with eigenvectors ${\sim}e^{i k x}$, to which the formulation is trivially adapted.

The obvious discretization, with $x_{\alpha} = -a + \alpha\Delta$, $\phi(x_{\alpha}) \mapsto \phi(\alpha)$, $\alpha = 1,\ldots, N=2a/\Delta$, then leads to (\ref{eq:eqmot1}), with $M(\alpha,\beta) = \rho(x_{\alpha}) \delta(\alpha,\beta)$, $\Gamma(\alpha, \beta) = \Gamma(x_{\alpha}) \delta(\alpha,\beta)$, and
\beq
  K(\alpha,\beta) = q^2 \delta(\alpha,\beta)
  -\frac{1}{\Delta^2} [ \delta(\alpha {-} 1, \beta) - 2
  \delta(\alpha,\beta) + \delta(\alpha {+} 1,\beta) ] \, .
\eeql{eq:cont05}
By reversing this mapping, it is straightforward to derive
the bilinear map in the continuum model, namely (up to an
irrelevant overall factor of $\Delta$)
\beq
  ( \bpsi , \bphi ) = i \int_{-a}^a dx \, \left[ {\hat \psi}(x)
  \phi(x) + \psi(x) {\hat \phi}(x) + \psi(x) \Gamma(x) \phi(x)
  \right] \quad .
\eeql{eq:cont06}

Examples with a $J=2$ critical point on the imaginary axis are trivial to construct. Take $\Gamma(x) = 2  \rho(x) \gamma $, with $\gamma$ to be tuned.  The system with $\gamma=0$ has a complete set of eigenfunctions: $ Kf_j = \Om_j^2 \rho f_j $, where $\Om_1^2< \Om_2^2 < \cdots$ are real. It then follows that $f_j$ are also eigenfunctions of (\ref{eq:cont03}), but with complex eigenvalues $ \om_j = -i\gamma \pm \sqrt{ \Om_j^2 - \gamma^2 }$. So as $\gamma$ is increased from zero, the modes $j = 1, 2, \ldots$ go through criticality in turn.

Examples with a critical point off the imaginary axis require that two parameters be tuned. Take $\Gamma(x) = 2 \rho(x)\gamma(x)$, with $\rho(x) = \rho_1$  ($\rho_2$) and $\gamma(x) = \gamma_1$ ($\gamma_2$) for $x < a_1$ ($x > a_1$), and $a=2$, $a_1=1$. Fix $q^2 =1$, $\rho_2=1$, $\gamma_2=1$ and tune $\rho_1$, $\gamma_1$. A critical point is found at $\rho_{1*} = 5.891$, $\gamma_{1*}=1.994$. One can again study perturbations of this critical point, say, $\rho_1 \mapsto \rho_{1*}+\ep$. The PF is evaluated using the bilinear map (\ref{eq:cont06}) and the norm $N_j = \int_{-a}^a dx\, \rho(x) |f_j(x)|^2$. We have verified that $|C_j| \propto\ep^{-1/2}$ for the near-critical modes (details not shown). In particular, models with very large values of $C_j$ are readily constructed.


\section{Nearly degenerate and weakly damped modes}
\label{app:b}

In this Appendix we consider some general properties of a $J{=}2$ Jordan block at $\om_* = \Om -i\delta$, where $\delta \ll \Om$ (weak damping) is regarded as a parameter.  This block is split by a small amount $\lambda$ (which may have any complex phase). The conjugate block at $-\Om -i\delta$ can be ignored for the present purpose. A crucial issue is that for $\delta \neq 0$, the 2-dimensional Jordan block contains only \emph{one} eigenvector (criticality), but for $\delta = 0$, the conservative subsystem does not allow criticality, so there must be be \emph{two} eigenvectors (level crossing). The $\delta \rightarrow 0$ limit is therefore subtle.

We operate only in the relevant subspace, assuming $\H = \H_0+\delta \H_1+ \cdots$, $K = K_0 + \delta K_1 + \cdots$, $M = M_0 +\delta M_1 + \cdots$ and $\Gamma = \delta \Gamma_1 + \cdots$, where importantly the damping matrix has no zero-order term. From (\ref{eq:cor06}), (\ref{eq:defbj}) and the definition of the PF in (\ref{eq:pet1}), correlation functions are sums of lorentzians, with amplitudes~$B_j$. Using the normalization (\ref{eq:jborth}) for the Jordan normal basis [specifically $(\g_0, \g_1) =1$], one has $\Nj \approx \pm 2 \lambda$ [cf.~(\ref{eq:normcrit})]. Near criticality, $\f_j \approx \g_0$, and we assume a series expansion in~$\delta$:
\beq
  \g_0=\delta^{\alpha} \left( \h_0 + \delta \h^{\prime}_0 +
  \cdots \right) \quad ,\qquad
  \g_1=\delta^{\beta} \left( \h_1 + \delta \h^{\prime}_1 +
  \cdots \right) \quad .
\eeql{eq:b03}
Now both $\g_0$ and $\g_1$ lie in the subspace $S(\delta)$ which is annihilated by $(\H-\om_*)^2$; for $\delta \rightarrow0$, they lie in $S(0)$, annihilated by $(\H_0-\Om)^2$. But for $\delta=0$, this subsystem is conservative, not allowing any criticality, so that these vectors must be annihilated by \emph{one} power of $\H_0-\Om$:
\beq
  (\H_0 - \Om) \h_0 = (\H_0 - \Om) \h_1 = 0 \quad .
\eeql{eq:b04}
The normalization condition $(\g_m , \g_m) = 0$ $\forall\,\delta$ implies for the leading terms $(\h_m , \h_m) = 0$, $m=0,1$.  This places strong restrictions on $\h_0$ and $\h_1$, which we do not spell out here, except to note
\beq
  0 = (\h_m , \h_m) = 2i \, h_m^{\rm T} \hat{h}_m^{\vphantom{\mathrm{T}}}
    = 2 \Om \, h_m^{\rm T} M h_m^{\vphantom{\mathrm{T}}} \quad.
\eeql{eq:b05}
The last normalization condition $(\g_0, \g_1) =1$ then gives, to leading order, $1 = \delta^{\alpha+\beta} (\h_0, \h_1)$, so that $\beta = -\alpha$ and $(\h_0 , \h_1) = 1$. Finally, $(\H - \om_{*}) \g_1 = \g_0$ gives
\beq
  \left[ (\H_0{ -} \Om) + \delta (\H_1 {+} i) + \cdots \right]
  \, \left[ \delta^{-\alpha} \left( \h_1 + \delta \h^{\prime}_1 +
  \cdots \right) \right]
  =\delta^{\alpha} \left( \h_0 + \delta \h^{\prime}_0 + \cdots
  \right) \quad .
\eeql{eq:b08}
The leading term on the LHS is $\delta^{-\alpha} (\H_0{ -}
\Om)\h_1 = 0$ by (\ref{eq:b04}). The next-leading terms give
\beq
  \delta^{-\alpha+1} \left[  (\H_0{ -} \Om) \h^{\prime}_1 +
  (\H_1 {+} i) \h_1 \right] = \delta^{\alpha} \h_0 \quad ,
\eeql{eq:b09}
implying $\alpha = \frac{1}{2}$ generically\footnote{Consider $\p(\delta) \equiv \delta^{\alpha} (\H-\om_*) \g_1$. By (\ref{eq:b04}), $\p(0) = 0$. If $-\alpha+1 < \alpha$, then the LHS of (\ref{eq:b09}) must alone vanish, corresponding to $\p(\delta)$ having a second-order zero, a non-generic case which we ignore. If $-\alpha+1 > \alpha$, then the leading term in (\ref{eq:b08}) cannot be satisfied.} and determining the next coefficient $\h^{\prime}_1$, details of which we shall not pursue. In short, $\g_0 \sim \delta^{1/2}$, $\g_1 \sim \delta^{-1/2}$, which is the analytic manifestation of the singular limit in which criticality goes over to level crossing.

Using these, we then find
\beq
  C_j \approx \frac{\delta}{\pm 2 \lambda}\,  \left( 2 \Om \,
  h_0^{\dagger} M h_0^\phd \right) \sim \pm \frac{\delta}{\lambda}\quad ,
\eeql{eq:b10}
while
\beq
  \sigma_j^{\prime} = \frac{ h_0^\mathrm{T} X h_0^{\vphantom{\mathrm{T}}} } {h_0^{\dagger} M h_0^\phd } \quad ,
\eeql{eq:b11}
the same value for $j=1,2$ and generically of $O(1)$.  This
then leads to (\ref{eq:clim}).  However, in the special case $X=M$,
by (\ref{eq:b05}) the numerator in (\ref{eq:b11}) vanishes to
$O(\delta^0)$, so that $B_j \sim \delta^2/\lambda$.

\begin{figure}
\centering
\includegraphics{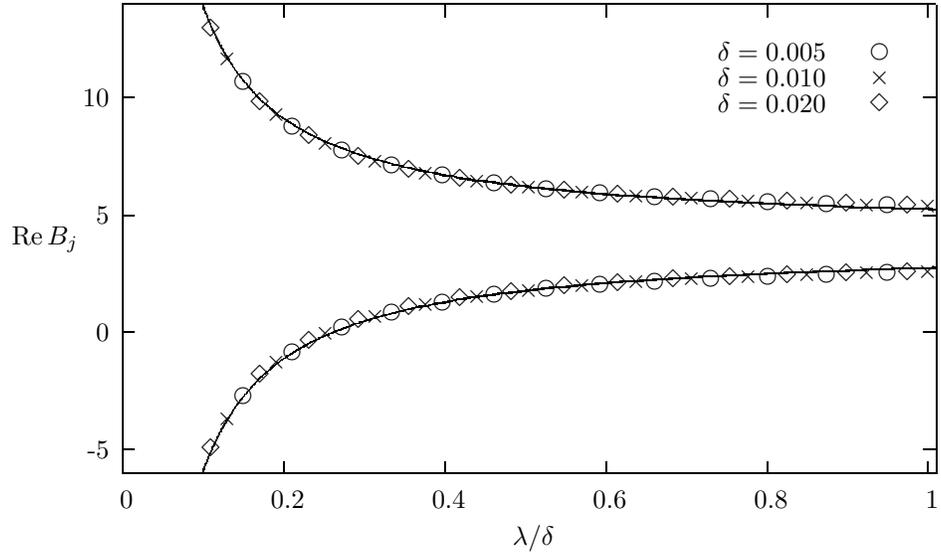}
\caption{$\Re B_j$ as a function of $\lambda/\delta$ for two
weakly damped and near-critical modes, for $\delta = 0.005$
(circles), $0.01$ (crosses), $0.02$ (diamonds). The curve is
(\ref{eq:bfit}).}
\label{fig4}
\end{figure}

The above analysis applies for $\lambda\ll\delta$. But for $\lambda \gg \delta$, one has two separate modes which are weakly damped; then by the usual arguments $C_{1,2} =1$. Again $\sigma_j^{\prime} = O(1)$, but in general having different values for $j = 1,2$. Thus one expects that $B_j$ in a broad range of $\lambda/\delta$ should be well described by
\beq
  B_{1,2} = \pm b (\delta/\lambda) + b^{\prime}_{1,2} \quad ,
\eeql{eq:bfit}
and this fit is shown by the line in Figure~\ref{fig4}.

\end{appendix}


\end{document}